\documentclass[12pt]{article}

\usepackage{amsthm,extmath}
\usepackage{latexsym}

\newcommand{\be}{\begin{equation}}
\newcommand{\ee}{\end{equation}}
\newcommand{\ba}{\begin{eqnarray}}
\newcommand{\ea}{\end{eqnarray}}

\theoremstyle{plain}
\newtheorem{tm}{Theorem}
\newtheorem{pr}[tm]{Proposition}

\theoremstyle{remark}
\newtheorem{rk}{Remark}
\newtheorem{example}[rk]{Example}

\theoremstyle{definition}
\newtheorem{df}{Definition}

\title{\bf Integral equation of quantum stochastic process }

\author{ {\bf Jerzy Stry\l{}a}
\footnote {e-mail address: jstryla@wp.pl }     }
\date{}

\begin{document}
\def\thesection{\Roman{section}}

\maketitle

\begin{abstract}
To describe stochastic quantum processes I propose an integral equation of
Volterra type
which is not generally transformable to any differential one. The process
is a composition of ordinary quantum evolution which admits presence of a
quantum bath and reductions to pure states. It is proved that generically
solutions stabilize asymptotically for $t\rightarrow +\infty$ to a
universal limit - the projection onto the state with maximal available
entropy. A number of typical methods of finding solutions of the equation
are proposed.
\end{abstract}

\newpage

\section{Introduction}

In classical evolution of a system-thermal bath it is observed existing of
time arrow. In quantum case there are a number of attempts to modify
equations to gain non-invertability of the dynamics. A basis for a formal
stochastic disturbance is usually a differential equation as the
generalized Schr\"odinger equation$^1$ or Lindblat equation, see for
example $^2$. The first modification leading to Hugstone equation is a
model of single measurement process. Nevertheless, it realizes one
reduction event as a long-time limit and the method is not applicable here,
where I assume that reductions moments are points on the time axis. In the
second approach the basis Lindblat equation is only a constant-coefficients
approximation of complete quantum evolution valid for relatively short time
regime, compare $^3$. In $^4$ it is described a quantum stochastic process
which avoids formalizing to differential equations at all. It consists from
mixed quantum evolutions and reductions, where it is assumed that the
reduction skips arise in time moments treated as a Poisson stochastic
process with a characteristic parameter $\nu $.

In section \ref{c1} I propose the new equation governing the evolution of
this type, which for $\nu = 0$ is equivalent to the mere quantum equation
(\ref{mere}). In section \ref{c2} the asymptotic theorem is proved.
Solutions of some special versions of the equation are investigated in
section \ref{solving}. In the appendix fulfilling of some genericity
conditions for the integral equations based on hamiltonian evolution are
elaborated.

\section{Integral equation}   \label{c1}
In the article I restrict myself to finite dimensional case. Let a quantum
evolution of a system in a system-bath pair is defined by
\ba
A_{\alpha}:{\bbold R}\rightarrow B(\cal{H}) \\
\sum_{\alpha}A_{\alpha}A^{\dagger}_{\alpha}=
\sum_{\alpha}A_{\alpha}^{\dagger}A_{\alpha}=1
\ea
where ${\cal H}={\bbold C}^n $, $B({\cal H})$ is the linear, bounded
operators set on ${\cal H}$, $\alpha =1,\ldots,N$, compare also with $^4\
{^3}$. Now, let $0=t_0<\ldots <t_i<t_{i+1}<\ldots $ be a realization of a
Poison process with the mean number of events in unit intervals equals to
$\nu\ge 0$. Between two neighboring points in which quantum reductions take
place $t_i,t_{i+1}$, the evolution is defined by the doubly stochastic
matrix ($2\Sigma $)
\ba     \label{mere}
M_{ij}(t):=\sum_{\alpha}TrP_{i}A_{\alpha}(t)P_jA_{\alpha}^{\dagger}(t)
\\ M_{ij}(0)=1,
\ea
where $t:= t_{i+1}-t_i$, an orthonormal basis $|i>\in {\cal H}, i=1,...,n$
is chosen and $P_i:= |i><i|$. As one parameter unitary group evolution is
determined by a self-adjoint operator in quantum bath presence case it is
determined by a set of operators $B_{ab}=B_{ba}^{\dagger }$. $A_{ab}(t)$,
$\alpha\equiv ab$, are solutions of
\be         \label{A}
\dot{A}_{ab}(t)=-i\sum_c B_{ac}A_{cb}(t)=-i\sum_c A_{ac}(t)B_{cb}
\ee
with the initial condition $A _{ab}(0)=\frac{\delta _{ab}}{\sqrt{n_2}}1$,
where $n_2$ is the dimension of the bath Hilbert space, $a,b,c = 1,2,\ldots
, n_2$.

Let $\bar{M}(T) $ be the average doubly matrix obtaining by summing up all
realizations of the Poisson process. The result is:
\ba   \nonumber
\bar{M}(T)=\sum_{n=0}^{\infty }\int_0^Tdt_n\int_0^{t_n}dt_{n-1} \ldots
\int_{0}^{t_2}dt_1M(T-t_n)\circ \ldots \\
\circ M(t_2-t_1)\circ M(t_1)\nu ^ne^{-\nu T}.     \label{M}
\ea
Following a structural similitude to Weinberg-Van Winter equation$^5$ one
obtains from (\ref{M}) the integral equation:
\be      \label{I}
\bar{M}(t)=e^{-\nu T}(M(T)+\nu \int_0^{T}M(T-t)\bar{M}(t)e^{\nu t}dt).
\ee
In the way I have built a dynamical system entirely defined by $B_{ab}\in
B({\cal H})$ and $\nu \ge 0$ by equations (\ref{A}, \ref{I}). The functions
$A_{\alpha }(t)$ given by (\ref{A}) are analytical, so $M(t)$ as well. Then
performing $k$ differentiations of (\ref{I}) I gain:
\be
\bar{M}(T)^{(k)}= e^{-\nu T} [M^{(k)}(T)+ L_k(T)+\nu \int_0^{T}M(T-
t)\bar{M}(t)^{(k)}e^{\nu t}dt]
\ee
where $L_k$ are defined by the recurrence formula:
\be
L_{k+1}(T)= \nu M(T)\bar{M}^{(k)}(0)+(1-\nu ) L_k(T)-\nu M^{(k)}(T).
\ee
Generally, the solution of (\ref{I}) is always uniquely defined by $M(t)$,
an integrable map from ${\bbold R}$ to the set of general stochastic
matrices and given by the convergent series (\ref{M}).
\begin{rk}
It is appeared that if one takes a different distribution of reductions
events and still wants
to keep an analog of equation (\ref{I}) it needs to be again a Poissonian
distribution. Nevertheless, the integral equation generalizes itself to the
following one:
\be
\bar{M}(T)=a(T)M(T)+ \int_0^{T}M(T-t)\bar{M}(t) b(t,T) dt,
\ee
where $a,b$ are nonnegative, continuous functions such that
\be
a(T)+\int_0^Ta(t_1)b(t_1,T)dt_1+\int_0^T\int_0^{t_1}a(t_2)b(t_2,t_1)
b(t_1,T)dt_2dt_1 +\ldots =1
\ee
or equivalently
\be
\int_0^Tb(t,T) dt = -a(T) + 1.
\ee
At the moment still if $M(t)$ are stochastic matrices (or $2\Sigma $) then
$\bar{M}(t)$ are of the same type.
\end{rk}

\section{Solving the equation}   \label{solving}
The following property of doubly stochastic matrices will be of interest.
\begin{df} \label{def}
For a doubly stochastic matrix $M:{\bbold R}^n\rightarrow {\bbold R}^n$ its
{\it compression} $c(M)$ is defined by $c(M):=||M|_{\triangle }||$, where
$M|_{\triangle}$ is the restriction of $M$ to the subspace of vectors $v^i$
fulfilling $\sum_{i=1}^{n}v^i=0$.
\end{df}
Clearly $0\le c(M)\le 1$, see $^4$. Now, I may consider some special situations
delivering more information about evident solutions constructing.

\begin{example}
The most simple one is $M(t)=M=const.$. Then equation (\ref{I}) is the
following:
\be
\dot{\bar{M}}=\nu (M-1)\bar{M}.
\ee
Then for $M=M^T$ the solution $\exp (\nu (M-1)t)$ has the limit for
$t\rightarrow \infty $ of the form
\be
\pmatrix{1 & 0 \cr
               0  &  \Theta  }.
\ee
In the case $c(M)<1$ the limit is $\Theta $, where $c(\Theta)=0 $.
\end{example}

\begin{example}                      \label{in-out}
Let $M(t)=\alpha (t)\cdot 1+(1-\alpha (t) )\cdot \Theta $, where $\alpha
,\beta $ are an integrable on finite intervals functions of values in the
interval $[0,1]$. Then equation (\ref{I}) is reduced to
\be                               \label{in-out eq}
\beta (T)=e^{-\nu T}\alpha (T)+\nu e^{-\nu T}\int_0^{T}
\alpha(T-t) \beta(t)e^{\nu t}dt,
\ee
where $\beta $ is uniquely defined through $\bar{M}(t) =\beta (t)\cdot 1+
(1-\beta (t)) \cdot \Theta $. In the similar way for each $M_1, M_2\in
2\Sigma $ such that $\{ \alpha M_1+(1-\alpha )M_2, \alpha \in [0,1] \}$ is
closed for matrices multiplication linear integral equations arise of the
form $\beta =\Omega _{\alpha } \beta$, where $\Omega _{\alpha }$ transforms
any integrable input function $\beta $ of values in $[0,1]$ into an output
$\Omega _{\alpha } \beta (t)\in [0,1]$. The solutions of the equations
(generically) stabilize in infinity to a number from $[0,1]$ as one may
conclude from the asymptotic theorem in section \ref{c2}.
\end{example}

In the case of locally constant functions $\alpha (t)$ except finite number
of discontinuities in bounded intervals equation (\ref{in-out eq}) may be
viewed as a sequences of differential equations defining and solving step
by step. I consider a simple $0, 1$ input function.

\begin{example}   \label{1010}
Let $\alpha (t) =\alpha_{2k}=1$ for $t\in [2k\tau , (2k+1)\tau ) $ and
$\alpha (t)=\alpha_{2k+1}=0$ for $t\in [(2k+1)\tau ,(2k+2)\tau )$, where
$k\in {\bbold N}$ and $\tau >0$. Then (\ref {in-out eq}) is transformable
to
\be
\dot{\beta}(T)=\nu \sum_{k=1}^i \beta(T-k\tau )
e^{-\nu k \tau }(\alpha _k-\alpha_{k-1})
\ee
with $T\in [i\tau ,(i+1)\tau )$ or adopting periodicity
\be   \label{periodic}
\dot{\beta}(T+2\tau )=e^{-2\nu \tau }\dot{\beta}(T)-\nu \beta(T+\tau)
e^{-\nu \tau} +\nu \beta (T)e^{-2\nu \tau}
\ee
The initial conditions for each intervals are $\beta (0)=\alpha (0)$ and
the discontinuity in $t=k\tau$ are given by $\beta (k\tau ^+) - \beta(k\tau
^-)= (-1)^{k}\exp(-\nu k\tau ) $ for $k\ge 1$. In the way solving the
integral equation in an interval one needs to possess already solutions of
last two. First two intervals need to be solved independently. Here $\beta
(T)=1$ for $T\in (0,\tau) $ and $\beta (T)= 1+ (\nu \tau -\nu T -1)
\exp(-\nu
\tau ) $ in $[\tau , 2\tau )$.
\end{example}

For effective finding of solutions I return to an analytical input.
\begin{example}   \label{cos}
Let
\be    \label{alpha}
\alpha (t) =\frac{1}{2}+\frac{\cos(t)}{2}=\frac{1}{2}+\frac{e^{it}}{4}+
\frac{e^{-it}}{4}
\ee
and $\nu =1$. I assume that the solution of (\ref{in-out eq}) has a form
\be
e^{t}\beta(t) = a(t) +b(t)e^{it} + \bar{b}(t) e^{-it}
\ee
where $a,b$ smooth functions, $a(t)\in {\bbold R}$, $b(t)\in {\bbold C}$.
Then the linear, constant coefficient equations follow
\ba
a^{(3)}-\ddot{a}+\dot{a}-\frac{1}{2}a=0, \\
b^{(3)}+(3i-1)\ddot{b}-2(i+1)\dot{b}-\frac{1}{4}b=0
\ea
with the initial conditions:
\ba
a(0)=\dot{a}(0)=\ddot{a}(0)=\frac{1}{2}, \\ b(0)=\frac{1}{4}, \\
\dot{b}(0)=-\frac{1}{4}, \\
\ddot{b}(0)=\frac{1}{4} (i-1).
\ea
Similarly, the equivalent, linear, constant coefficients differential system
of equations of order $2N+1$ may be constructed for $M(t)=\sum_{n=-N}^N A_n
\exp (int)$, where $A_n$ are constant matrices.
\end{example}

\section{Asymptotic theorem}  \label{c2}
I define a restriction on $2\Sigma^{ {\bbold R}_+} $ maps
\begin{df}    \label{generic}
Let $M:{\bbold R}_+\rightarrow M(n\times n,{\bbold R}) $ be a measurable
map into doubly stochastic matrices. It is said to be {\it generic } iff
$\exists ( \delta < 1 )\ \mu (I_{\delta })>0 $, where $I_{\delta }:
=\{ t \in {\bbold R}_+; c(M(t)) \le \delta \}$ and $\mu $ the Lebesgue measure of
the real, non-negative numbers space ${\bbold R}_+$.
\end{df}
The above genericity condition naturally appears for $M(t)$ generated by
physical quantum systems defined in section \ref{c1}, for details see
appendix \ref{c3}. Now, the main theorem may be proved.

\begin{tm}    \label{main}
Let $M(t)$ be a generic map. Then $\lim_{t\rightarrow
\infty}\bar{M}(t)=\Theta $.
\end{tm}
\begin{proof}
One has the following estimation

\ba   \nonumber
c(\bar{M}(t))\le \sum_{n=0}^{\infty }\int_0^Tdt_n\int_0^{t_n}dt_{n-1}
\ldots
\int_{0}^{t_2}dt_1c(M(T-t_n))\circ \ldots      \\
\circ c(M(t_2-t_1))\circ c(M(t_1))\nu ^ne^{-\nu T}.   \label{cM}
\ea
so the problem is reduced to considering equation (\ref{in-out eq}) with
$\alpha $ differing from $1$ on a positive measure set. Changing variables
in the integral of the equation and making transformation $T\rightarrow
T/\nu $ lead to elimination of $\nu $ ($\nu=1$). Let
\be
\delta = \int_{0}^{\infty }\alpha(t) e^{-t}  dt.
\ee
From the assumption about $\alpha $ is that $\delta < 1$. Now, let
$\beta(T)\le b_0$ for $T\in [t_0, \infty)$. From (\ref{in-out eq}) I have
\be     \label{est}
\beta (T)\le e^{-T}\alpha (T)+e^{-T}\int_0^{t_0}\alpha(T-t) \beta (t) e^{t} dt +
b_0 \delta .
\ee
I define $\epsilon = (1-\delta)/2$. Let $t_0'$, $t_0'\ge t_0$, be such that
two first terms of (\ref{est}) are less or equal to $b_0\epsilon $ for $t\ge
t_0'$. Then $\beta (T)\le b_0 (\delta + \epsilon )$ for $T\ge t_0'$.
Continuing the procedure one reaches $\lim_{T\rightarrow \infty}
\beta(T)=0$.
\end{proof}

Nevertheless, even for $c(M(t))=1$ an asymptotic stabilization can exist as
the example is showing.
\begin{example}
Let $M(t)=P $, where $P$ is a cyclic group generator such that $P^i\ne 1$
for $i=1,\ldots ,k-1 $ and $P^k =1$. I also assume that $c(\sum_{i=1}^{k}
\alpha _i P^i)=1$ for $\alpha_i\ge 0$ and $\sum_{i=1}^{k} \alpha_i =1$. It
may be easy realized by the matrix
\be
\pmatrix{P_{\sigma} &  0  \cr
                0                   &  P_{\sigma}   }.
\ee
$P_{\sigma } $ is a permutation. Then the solution (\ref{M}) has the form
\be
\bar{M}(t) = \exp(-\nu t)[P^kf_1(\nu t) + Pf_2(\nu t)
  + \ldots +P^{k-1}f_k(\nu t)] P,
\ee
where the analytical functions, $f_i(z)$, are formally defined by
$\exp(\alpha_i z)=(\alpha_i)^k f_1(z)+\ldots +(\alpha_i)^{k-1} f_k(z)$ with
$\alpha_i^k=1$. Then $\lim_{t\rightarrow \infty }\bar{M}(t)
=\frac{1}{k} \sum_{i=1}^k P^k$.
\end{example}

Therefore, a generalization may be proposed.
\begin{tm}   \label{gener}
Let $M\in 2\Sigma^{\bbold{R}_+}$ be a right-side continuous map and
$M(0)=1$. If $\bar{M}(t)$ fulfills the integral equation then
\be  \label{lim}
\lim_{t\rightarrow \infty}\bar{M}(t)=
                                  \pmatrix{\Theta_1 & 0 &  &0 &0 \cr
                                             0 & \Theta_2  &  & 0 &0  \cr
                                              &    & \ddots &    &      \cr
                                             0&0&   &  \Theta_N&0  \cr
                                             0 & 0 &   & 0  & id  }.
\ee
\end{tm}
\begin{proof}
I put $S_M:=\{M(t);t\in {\bbold R}_+ \}$.
Let $\bar{S}_M$ be the closure of $S_M$ with respect to matrices
multiplication and $\frak{I}:=\{ P_{\sigma}; \sigma \in I \}$ be a
minimal permutations' subset spanning $\bar{S}_M$. Then ${\bbold
R}^n=\bigoplus_i V_i\oplus V_{id}$, where $V_i\ne \{0\} $ are minimal
invariant subspaces of $P\circ A_{\frak{I}}$, $P^{-1}\in \frak{I}$,
spanned by the canonical basis vectors and such that $\dim V_i>1$, where
$A_{\frak{I}}$ denotes all doubly stochastic
matrices obtained from $\frak{I}$ as baricentric points,
see proposition \ref{decomp}.
On $V_{id}$, $V_{id}\perp V_i$,
elements of $A_{\frak{I}}$ acts as
identity. $1\in A_{\frak{I}}$, so $P=1$ may be chosen.
I define a compression in $V_i$ by $c_i(M):=c(M|_{V_i})$ for
$M\in A_{\frak{I}}$. If $\forall(M\in A_{\frak{I}})\ c_i(M)=1$ then
$P'\circ M|_{V_i}$ is decomposable for $P'$ being an admitted permutation
of the basis vectors from $V_i$.
Therefore, for all interior $M\in A_{\frak{I}}^i$ the compression
$c_i(M)<1$, also the next consequence
is that $\Theta_i\in A_{\frak{I}}^i$,
where I have denoted $A_{\frak{I}}^i:=A_{\frak{I}}|_{V_i}$.
Now, I restrict considerations to one subspace $V_i$ and
for simplicity omit the index.
Let $\frak{S}\subset \bar{S}_M $ be a minimal set that does not belong
to any $A_{\frak{J}}$, $\frak{J}\varsubsetneq \frak{I}$. $\frak{S}$ is finite
and $c(\sum_k \alpha_k \frak{s}_k)<1$ for $\alpha_k>0$, $\sum_k \alpha_k
=1$ and $\{\frak{s}_k \}=\frak{S}$. Each $\frak{s}_k$ has a form
$M(t_{N_k})\circ\ldots \circ M(t_{1_k} )$. Let $N:=\max \{N_k \}$. At the
moment I return to equation (\ref{I}) with $\nu =1$. Equivalently, it may be
written in the form:
\ba  \label{IN}
\bar{M}(T)=e^{-T}M(T)+e^{-T}\int_0^T M(T-t_1)M(t_1)dt_1+\ldots  \\
\nonumber
+e^{-T} \int_0^T \int_0^{t_N}\ldots\int_0^{t_2}M(T-t_N)\circ \ldots
\circ M(t_2-t_1)\bar{M}(t_1)e^{t_1}dt_1\ldots dt_N.
\ea
I will check the limit of $W(T):=\sum_l \gamma_l \bar{M}(T+T_l)$, where
$\gamma_l>0 $, $\sum_l \gamma_l=1$ and
\be
T_k:=\sum_{n_k=1}^{N_k} t_{n_k},
\ee
$k$ are indices of $\frak{s}_k$.

The final step is like as in theorem \ref{main}. Having an upper bound $b_0$
of $c(W(t))$ in $[t_0,\infty )$ one improves it, here using (\ref{IN}),
to $(1+\delta) b_0/2$ on $[t_0',
\infty )$ in the next step, where $t_0'\ge t_0+1$ may be found.
Now, I will show it. The matrix $W(T)$ has the form:
\ba   \nonumber
W(T)=\sum_l \gamma_l \Bigl( e^{-(T+T_l)} M(T+T_l) +\int_0^{T+T_l}M(T+
T_l-t_1) M(t_1) dt_1+\ldots  \Bigr)  \\  \nonumber
+\sum_l \gamma_l e^{-(T+T_l)} \int_0^{T+T_l} \int_0^{t_N}\ldots\int_0^{t_0}
M(T+T_l-t_N)\circ \ldots \\ \nonumber
\circ M(t_2-t_1)\bar{M}(t_1)e^{t_1}dt_1\ldots
dt_N+ \\ \label{W(t)}
\sum_l \gamma_l e^{-(T+T_l)} \int_0^{T+T_l} \int_0^{t_N}\ldots\int_{t_0}^{t_2}
M(T+T_l-t_N)\circ \ldots \\  \nonumber
\circ M(t_2-t_1)\bar{M}(t_1)e^{t_1}dt_1\ldots dt_N.
\ea
One may verify that
\ba  \nonumber
c \Bigl( \sum_l \gamma_l \int_{\Delta }M(t_{N_l}-\epsilon_N)\circ\ldots %
\circ M(t_{1_l}+\epsilon_2-\epsilon_1)
\bar{M}(T+\epsilon_1 ) e^{-T_l+\epsilon_1} d^N \epsilon \Bigr) \\
< \sum_l  \gamma_l e^{-T_l}\mu(\Delta ) \sup_{\Delta } c(\bar{M}(T+
\epsilon_1)),
\ea
where $\Delta $ is a measurable set of $\bbold{R}^N_+$, $\mu(\Delta)>0$
and $\frak{s}_l=M(t_{N_l})\circ \ldots \circ
M(t_{1_l})$ with additional $M(0)=1$ if it is necessary for $N_l=N$.
In the way I obtain an estimation for the last term of
(\ref{W(t)}) by $\delta b_0$, $\delta<1$.
The first terms of (\ref{W(t)}) have vanishing  compression for
$T\rightarrow \infty $. The proof is completed.
\end{proof}
In other words the limit doubly stochastic matrix projects initial state
$p$ onto the maximal entropy state available for $p$.
Theorem \ref{gener} covers all analytical situations arising from quantum
physics. Nevertheless, for keeping the
result (\ref{lim}) it is also enough to assume that $M(t)$ differs from
maps in theorem \ref{gener} on a set of measure zero.

\section{Summary and interpretation}
I have modified the ordinary quantum evolution of a system-bath through
considering an associated stochastic process describing probability
transformation, not Hilbert space vectors. The arising formalism offers an
integral equation which transgresses the former methods of stochastic
modification of differential equation based on adding a linear stochastic
term. The closer analysis of the integral equation direct
solving shows the
difference with the ordinary differential one, its non-Markovian character.
The equation appears in a natural way from physical considerations and
reproduces the time arrow - the missing feature of unitary evolution. The
relation to the physical origin is stressed by a simple observation. Let
$M(t)$ be of the period $2\pi$. If I change the period by a general $\tau >0$
then $\bar{M}'(t)$ corresponding to $M'(t)=M(\frac{2\pi}{\tau}t)$ is equal
to $\bar{M}(\frac{2\pi}{\tau}t)$, but, now, the integral equation is governed
by the new $\nu '$ such that
\be
\tau \nu' =2\pi \nu.
\ee
This is a theoretical suggestion for a universal relation between the
period of quantum wave and the associated coefficient of stochastic
reduction. The interpretation may be found even for cases excluded from the
asymptotic theorem \ref{main}. Namely, if $M(t)$ contains an identity
sector arising from eigenvectors of hamiltonian, then in the sector the
decay effect does not appear at all. In the way $\nu >0$ influences only
states which during evolution cease to be pure. Appearance of invariant
sectors is related to observables commuting with the hamiltonian,
which are not
also disturbed by the stochastic modification.

\appendix
\section{Genericity conditions}  \label{c3}
The aim of the appendix is to study the maps $M(t)$ from ${\bbold R}_+$
into $2\Sigma $ arising from quantum evolution. Firstly, I start from
a characterization of doubly stochastic matrices with unit compression.
\begin{pr}      \label{decomp}
Let $M$ be a doubly stochastic matrix. Then $c(M)=1$ iff $P\circ M$ is a
decomposable matrix for $P$ being a permutation.
\end{pr}
\begin{proof}
The implication ($\Leftarrow $) is obvious. If ${\bbold R}^n=V_1\oplus V_2$
is a decomposition then it is enough to put $p=p_1^e\oplus q p_2^e $, $q\ne
1$ to have $p\ne p^e$ and $||M(p)||=||P\circ M(p)||=||p||$, where
$(p^e)^k=1/\dim V$ and $(p^e_i)^k=1/\dim V_i$, $i=1,2$. To show
($\Rightarrow $) I take $M=\sum_{i} \alpha_i P_i $, where $\sum_i \alpha
_i=1$, $\alpha_i \ge 0$ and $P_i$ are permutations enumerated by $i$. I may
assume that $\alpha_1\ne 0 $. Then $P_1^{-1} \circ M=\alpha_1 1+
P_1^{-1}\circ \sum_{i\ne 1}\alpha_i P_i $. Still $c(P_1^{-1}\circ M)=1$, so
$p$, $p\in \triangle $, exists such that $ P_1^{-1}\circ P_i
p=p$ and  $||M(p)||=||p||$.
Then the canonical basis vectors for which corresponding components
$p^k$ are equal to themselves constitute the invariant subspaces and a
decomposition is done.
\end{proof}
From the above proposition one states that $c(M)<1$ is open and dense in
$2\Sigma $.

Doubly stochastic matrices can be also built via Kraus representation$^4$.
Then one begins from $\{ A_{\alpha } (t) \}_{\alpha =1}^N $, defined by
equation (\ref{A}). Generic maps $M(t)$ defined by (\ref{mere})
appear in the following way.
\begin{pr}
If at least one $B_{ab}'\in \{B_{ab} \}$ is not proportional to the
identity then a dense and open set of orthonormal basises of ${\bbold C}^n$
exists such that $M(A(t))$ is generic.
\end{pr}
\begin{proof}
Now, $M(t)=M(A(t))$ defined by an orthonormal basis $\{|i> \}$ appears to
be analytical, so of the form
\be
M(t)=\sum_{k=0}^{\infty} \frac{t^k}{k!} M_k,
\ee
where $M_k\in M(n\times n, {\bbold R})$. First terms of the expansion are
$M_0=1$, $M_1=0$ and
\be
(M_2)_{jl}= \frac{2}{n_2}
\Bigl( \sum_{a,b,c}|<j|B_{ab}|l>|^2-<j|B_{ac}B_{ca}|l>\delta_{jl} \Bigr).
\ee
Basises $\{ |i> \}$ constituting the open and dense set of $U(n)$ are such
that $j\ne l \ \Rightarrow \ <j|B_{ab}'|l>\ne 0$, where $B_{ab}'$ is
indicated in the assumption. One finds that
\be   \label{gen}
||M(t)p||=1+\sum_{i,j} p_i(M_2)_{ij}p_j t^2/2+o(t^2) ,
\ee
where $||p||=1$. Further
\be
\sum_{i,j} p_i(M_2)_{ij} p_j=\sum_{i,j} \sum_{a,b}|<i|B_{ab}|j>|^2\Bigl( p_ip_j
-\frac{p_i^2+p_j^2 }{2}\Bigr) \le 0.
\ee
Therefore, for a generic basis $\{ |i> \}$ from (\ref{gen}) one obtains
$c(M(t))<1$ for a $t>0$. The genericity of $M(t)$ is proved.
\end{proof}

\newpage

\noindent
$^1$ D. C. Brody and P. Hughston, e-preprint, quant-ph/0011125 \\ \\ \\ \\
$^2$ H. M. Wiseman and L. Di\'osi, e-preprint, quant-ph/0012016 \\ \\ \\ \\
$^3$ D. A. Lindar and Z. Bihary and K. B. Whaley, e-preprint,
cond-math/0011204 \\ \\ \\ \\ $^4$ J. Stry\l{}a, Quantum stochastic
process, (2001), to appear in Acta Phys. Polonica B
\\ \\ \\ \\ $^5$ W. Thirring, {\it A Course in
Mathematical Physics}, (Springer-Verlag, New York, 1981) vol. 3 \\ \\



\end{document}